\documentclass[notitlepage,a4paper,aps,prd,onecolumn,superscriptaddress,nofootinbib,groupedaddress,longbibliography]{revtex4-1}
\usepackage{amsmath}
\usepackage{amsfonts}
\usepackage{amssymb}
\usepackage[utf8]{inputenc}
\usepackage[T1]{fontenc}
\usepackage[colorlinks=true]{hyperref}
\usepackage[dvipsnames]{xcolor}
\usepackage{graphicx}
\usepackage[normalem]{ulem}

\newcommand{\uu}[0]{\mathrm{u}}
\newcommand{\vv}[0]{\mathrm{v}}
\newcommand{\ww}[0]{\mathrm{w}}
\newcommand{\xx}[0]{\mathrm{x}}
\newcommand{\D}[0]{\mathrm{D}}
\newcommand{\dd}[0]{\mathrm{d}}

\begin{document}

\title{Propagation of gravitational waves in teleparallel gravity theories}

\author{Manuel Hohmann}
\email{manuel.hohmann@ut.ee}
\affiliation{Laboratory of Theoretical Physics, Institute of Physics, University of Tartu, W. Ostwaldi 1, 50411 Tartu, Estonia}

\author{Martin Kr\v{s}\v{s}\'ak}
\email{martin.krssak@ut.ee}
\affiliation{Laboratory of Theoretical Physics, Institute of Physics, University of Tartu, W. Ostwaldi 1, 50411 Tartu, Estonia}

\author{Christian Pfeifer}
\email{christian.pfeifer@ut.ee}
\affiliation{Laboratory of Theoretical Physics, Institute of Physics, University of Tartu, W. Ostwaldi 1, 50411 Tartu, Estonia}

\author{Ulbossyn Ualikhanova}
\email{ulbossyn.ualikhanova@ut.ee}
\affiliation{Laboratory of Theoretical Physics, Institute of Physics, University of Tartu, W. Ostwaldi 1, 50411 Tartu, Estonia}

\begin{abstract}
We investigate the propagation of gravitational waves in the most general teleparallel gravity model with second order field equations as perturbations around the Minkowski background. We argue that in this case the most general Lagrangian at the first non-vanishing order of the perturbations is given by a linear combination of quadratic invariants and hence coincides with the well-known new general relativity model. We derive the linearized field equations and analyze them using the principal polynomial and the Newman-Penrose formalism. We demonstrate that all gravitational wave modes propagate at the speed of light and there are up to six possible polarizations. We show that two tensorial modes of general relativity are always present and the number of extra polarizations depends on the free parameters of the new general relativity model.
\end{abstract}

\maketitle

\section{Introduction}\label{sec:intro}
Modified gravity theories are a viable alternative to dark energy in addressing the problem of accelerated expansion of the Universe \cite{Nojiri:2006ri,Capozziello:2011et}. 
A novel class of modified gravity models that caught a lot of attention recently are the so-called \textit{modified teleparallel theories}. These theories are motivated by the fact that the ordinary general relativity (GR) can be reformulated using the teleparallel geometry resulting in a theory known as the \textit{teleparallel equivalent of general relativity} (TEGR) or shortly just \textit{teleparallel gravity} 
\cite{Einstein1928,Sauer:2004hj,Moller1961,Hayashi:1967se,Cho:1975dh,Hayashi:1977jd,AP,Maluf:2013gaa}.

While TEGR is equivalent to the ordinary formulation of GR in terms of curvature in all physical predictions
, this equivalence is lost when we consider modified gravity theories based on these different underlying geometries. The most well-known example is the case of $f(T)$ gravity, constructed in analogy with $f(R)$ gravity, where the Lagrangian is taken to be an arbitrary function of the so-called torsion scalar, which defines the TEGR action \cite{Ferraro:2006jd,Ferraro:2008ey,Bengochea:2008gz,Linder:2010py}. When a non-linear function $f$ is considered, the resulting $f(T)$ theory represents a novel gravity model with rich dynamics distinctive from $f(R)$ gravity. See \cite{Cai:2015emx} for an extensive overview.

The recent discovery of gravitational waves \cite{Abbott:2016blz,Abbott:2017oio} opened a new way to test various modified theories of gravity~\cite{Lombriser:2015sxa,Lombriser:2016yzn,Chakraborty:2017qve,Sakstein:2017xjx,Ezquiaga:2017ekz,Baker:2017hug,Akrami:2018yjz}. This motivates a study of gravitational waves in modified gravity theories and proper understanding of their fundamental properties. Particularly interesting are the questions about the number of polarization modes of gravitational waves and their corresponding propagation velocities. The case of $f(R)$ gravity is well-understood and it has been shown that these theories all possess an additional massive scalar gravitational wave mode
\cite{Chiba:2003ir,Corda:2007hi,Naf:2011za,Yang:2011cp} compared to GR.

In the case of modified teleparallel theories, gravitational waves have been studied first in the case of $f(T)$ gravity \cite{Bamba:2013ooa,Cai:2018rzd}, where, in contrast to the $f(R)$ case, it was shown there are no extra propagating gravitational modes compared to GR. As we will argue later, this follows from a simple observation that $f(T)$ gravity effectively reduces to TEGR at the perturbative level and hence we obtain only the usual two GR polarizations. Only very recently \cite{Abedi:2017jqx,Farrugia:2018gyz}, it was shown that new polarization modes appear if we extend $f(T)$ gravity by introducing scalar fields or higher-derivative terms of the torsion in the case of so called $f(T,B)$ \cite{Bahamonde:2015zma} and $f(T,T_G)$ \cite{Kofinas:2014owa} theories, where $B$ is the boundary term relating the Riemannian curvature scalar with the torsion scalar and $T_G$ is the teleparallel equivalent of the Gauss-Bonnet term.

In this paper we follow another approach and study gravitational waves propagating around the Minkowski background in the model known as \textit{new general relativity} (NGR) \cite{PhysRevD.19.3524}, where the Lagrangian is taken to be a most general linear combination of quadratic parity preserving torsion invariants.\footnote{Note that sometimes "new general relativity" refers only to a special subclass of these theories in which  only one of the three parameters we consider here is left free and two are fixed to a specific value.} Our study is motivated by a simple observation that, unless we introduce higher derivatives or scalar fields, the most general teleparallel gravity Lagrangian at the perturbative level is given by the linear combination of quadratic invariants of the torsion, i.e., NGR. For example, recently proposed $f(T_\text{ax},T_\text{ten},T_\text{vec})$ gravity \cite{Bahamonde:2017wwk} and the so-called axiomatic electrodynamics inspired models \cite{Hohmann:2017duq}, which are both very general frameworks designed to include all previously studied teleparallel models as special instances, reduce to the case of NGR at the perturbative level.

We analyze the gravitational waves in the NGR model using two methods. Firstly, we consider a perturbative analysis of the NGR model following the example of \cite{Shirafuji:1996im,Obukhov:2002hy} and analyze the resulting linearized field equations by the method of the principal symbol. Secondly, we use the Newman-Penrose formalism \cite{Newman:1961qr} and classify the resulting polarizations according to the classification scheme introduced in~\cite{Eardley:1973br,Eardley:1974nw}. We show that all gravitational modes propagate at the speed of light and derive how the number of polarization modes depends on the free parameters of the NGR Lagrangian.

The outline of this paper is as follows. In section~\ref{sec:lintelep} we briefly introduce teleparallel geometry and the NGR model as the most general teleparallel gravity at the perturbative level. In section~\ref{sec:speed} we introduce the principal symbol and derive that all gravitational wave modes propagate at the speed of light. In section~\ref{sec:polar} we use the Newman-Penrose formalism to analyze the possible polarizations of gravitational waves and show how they depend on the free parameters of the NGR Lagrangian. We conclude this paper with a brief discussion and outlook in section~\ref{sec:conclusion}.

In this article we use the following notation. Latin letters $a,b,\ldots$ are Lorentz indices and Greek letters $\mu,\nu,\ldots$ are spacetime coordinate indices. The Minkowski metric is denoted by $\eta$ and has components $\eta_{ab} = \textrm{diag}(-1,1,1,1)$.

\section{Linearized Teleparallel Gravity}\label{sec:lintelep}
We begin this article with a short review of the required geometric notions in teleparallel gravity in section~\ref{ssec:telegeom}. We then recall the NGR Lagrangian in section~\ref{ssec:feq}, where we also derive the corresponding linearized field equations.

\subsection{Teleparallel geometry}\label{ssec:telegeom}
The fundamental variables in theories of gravity formulated in terms of teleparallelism are the tetrad $1$-forms $\theta^a$, their dual vector fields $e_a$ and the curvature free spin connection $\omega^a{}_b$ generated by local Lorentz transformations $\Lambda^a{}_b$. In local coordinates on spacetime they can be expressed as
\begin{align}
	\theta^a = \theta^a{}_\mu \dd x^\mu,\quad e_a = e_a{}^\mu \partial_\mu,\quad \omega^a{}_b(\Lambda) = \omega^a{}_{b\mu}(\Lambda) \dd x^\mu = \Lambda^a{}_q \dd (\Lambda^{-1})^q{}_b = \Lambda^a{}_q \partial_\mu (\Lambda^{-1})^q{}_b\dd x^\mu\,.
\end{align}
Moreover the tetrad $1$-forms and their duals satisfy
\begin{align}\label{eq:tetradinv}
	\theta^a(e_b) = \theta^a{}_\mu e_b{}^\mu = \delta^a_b,\quad \theta^a{}_\mu e_a{}^\nu = \delta^\nu_\mu\,,
\end{align}
and define a Lorentzian spacetime metric via
\begin{align}\label{eq:g}
	g_{\mu\nu} = \eta_{ab}\theta^a{}_\mu \theta^b{}_\nu,\quad g^{\mu\nu} = \eta^{ab}e_a{}^\mu e_b{}^\nu\,.
\end{align}
Tensor fields can be expressed either in coordinate or tetrad basis. For a $(1,1)$-tensor $Z$ we may for example write
\begin{align}
	Z = Z^\mu{}_\nu\ \dd x^\nu \otimes \partial_\mu = Z^a{}_b\ \theta^b \otimes e_a\,.
\end{align}
Thus when we change an index from Latin to Greek, this operation is done via multiplication with $\theta^a{}_\mu$ or $e_a{}^{\mu}$ respectively.

The building block of Lagrange densities is the torsion of the spin-connection given by
\begin{align}
	T^a = \D \theta^a = (\partial_\mu \theta^a{}_\nu + \omega^a{}_{b\mu}\theta^b{}_\nu) \dd x^\mu \wedge \dd x^\nu\,,
\end{align}
where the spin covariant derivative $\D$ ensures a covariant transformation behaviour under local Lorentz transformations of the tetrad \cite{Krssak:2015rqa,Krssak:2015oua}. More precisely, consider a tetrad $\hat \theta^a$ which is related to the original tetrad by a local Lorentz transformation $\tilde \Lambda^a{}_b$, i.e., $\hat \theta^a = \tilde \Lambda^a{}_b \theta^b$. Then, the torsion tensor of the tetrads are related by $\hat T^a = \tilde \Lambda^a{}_b T^b$, where the connections are given in terms of two further Lorentz transformations $\hat \Lambda$ and $\Lambda$
\begin{align}\label{eqn:flatspinconn}
	\hat \omega^a{}_b
	= \hat \Lambda^a{}_c \dd (\hat \Lambda^{-1})^c{}_b,\quad \omega^{a}{}_b
	= (\tilde \Lambda^{-1})^a{}_{c}\hat{\Lambda}^c{}_d \dd (\tilde \Lambda^e{}_b (\hat\Lambda^{-1})^d{}_e)
	= \Lambda^a{}_d \dd (\Lambda^{-1})^d{}_b\,.
\end{align}
In particular when one considers $\tilde \Lambda = \hat \Lambda$ one chooses the so-called proper tetrad or a tetrad in the Weitzenbock gauge, for which $\omega^a{}_b = 0$~\cite{Krssak:2015oua}.

The components of the torsion in local coordinates are therefore canonically labelled by $T^a = \frac{1}{2}T^a{}_{\mu\nu} \dd x^\mu \wedge \dd x^\nu$. In the following we will use the torsion components with spacetime indices only obtained via $T^\alpha{}_{\mu\nu} = T^a{}_{\mu\nu}e_a{}^\alpha$.

\subsection{Lagrange density and field equations}\label{ssec:feq}
We consider here the \emph{new general relativity} (NGR) \cite{PhysRevD.19.3524} model given by the action
\begin{align}\label{eq:Lag}
L_{\text{tot}}(\theta, \partial \theta, \Lambda, \partial \Lambda,\Phi^I)
&= 	L(\theta, \partial \theta, \Lambda, \partial \Lambda) + 	L_\text{M}(\theta,\Phi^I )
\end{align}
where $L_\text{M}(\theta,\Phi^I )$ is the matter Lagrangian, which is constructed via the usual minimal coupling principle. The spacetime metrics appearing during that procedure are understood as functions of the tetrads. The gravitational Lagrangian is the most general Lagrange density quadratic in the torsion tensor
\begin{align}\label{eq:LNGR}
	L(\theta, \partial \theta, \Lambda, \partial \Lambda)
	&= |\theta| \big(c_1 T^\rho{}_{\mu\nu}T_{\rho}{}^{\mu\nu} + c_2T^\rho{}_{\mu\nu} T^{\nu\mu}{}_\rho + c_3 T^\rho{}_{\mu\rho}T^{\sigma\mu}{}_{\sigma}\big)
	= |\theta| G_{\alpha\beta}{}^{\mu\nu\rho\sigma}T^{\alpha}{}_{\mu\nu}T^\beta{}_{\rho\sigma}\,,
\end{align}
where three real parameters $c_1, c_2$ and $c_3$ define different NGR theories.
In the last equality we introduced the supermetric \cite{Ferraro:2016wht} or constitutive tensor \cite{Itin:2016nxk,Hohmann:2017duq}
\begin{align}\label{eq:G}
	G_{\alpha\beta}{}^{\mu\nu\rho\sigma} =c_1 g_{\alpha\beta}g^{\rho[\mu}g^{\nu]\sigma} - c_2 \delta_\beta^{[\mu}g^{\nu][\rho}\delta^{\sigma]}_\alpha - c_3\delta_\alpha^{[\mu}g^{\nu][\rho}\delta^{\sigma]}_\beta\,,
\end{align}
which will turn out to be convenient for the following analysis. The appearing spacetime metric $g$ is understood as function of the tetrads \eqref{eq:g}. The teleparallel equivalent of general relativity (TEGR) is included in the NGR class of gravity theories for the choice $c_1 = \frac{1}{4}, c_2 =\frac{1}{2} $ and $c_3 =-1 $.

To analyse the propagation of gravitational waves for NGR gravity around the Minkowski background we derive the linearized field equations of the theory. To do so we fix Cartesian coordinates \((x^{\mu}, \mu = 0, \ldots, 3)\) and make the following perturbative ansatz for the tetrad and the Lorentz transformation defining the spin connection
\begin{subequations}\label{eqn:pertfields}
\begin{align}
\theta^a{}_\mu &= \delta^a_\mu + \varepsilon\ \uu^a{}_\mu\\
e_a{}^\mu &= \delta_a^\mu + \varepsilon\ \vv_a{}^\mu\\
\Lambda^a{}_b &= \delta^a_b + \varepsilon\ \ww^a{}_b\,,
\end{align}
\end{subequations}
where $\varepsilon$ is a perturbation parameter. The duality between $\theta^a$ and $e_a$ implies to first order in $\varepsilon$ that $\vv_a{}^\mu \delta_\mu^b = - \uu^b{}_\nu \delta^\nu_a$ and $\Lambda$ being a local Lorentz transformation implies that $\ww_{ab} = - \ww_{ba}$. The perturbative gauge transformations are
\begin{align}
\begin{split}\label{eq:pertgt}
	\hat \theta^a = \tilde \Lambda^a{}_b \theta^b &\Rightarrow \hat \uu^a{}_\mu = \tilde \ww^a{}_\mu + \uu^a{}_\mu\\
	\Lambda^a{}_d = (\tilde \Lambda^{-1})^a{}_{c}\hat{\Lambda}^c{}_d &\Rightarrow \hat \ww^a{}_b = \tilde \ww^a{}_b + \ww^a{}_b\,.
\end{split}
\end{align}
Moreover changing the index type from Lorentz to spacetime, to first order in the perturbation, is done with a $\delta^a{}_\mu$ resp. $\delta^\mu{}_a$ and raising and lowering any kind of index is done with the Minkowski metric $\eta_{ab}$ res. $\eta_{\mu\nu}$ or its inverse.

The torsion tensor can be expanded into the first order fields as
\begin{align}
	T^a{}_{\mu\nu} = 2 \partial_{[\mu} \theta^a{}_{\nu]} + 2\omega^a{}_{b[\mu}\theta^b{}_{\nu]} =2 \varepsilon \big( \partial_{[\mu} \uu^a{}_{\nu]} - \partial_{[\mu} \ww^a{}_{\nu]} \big) + \mathcal{O}(\varepsilon^2)\,.
\end{align}
In this order of the perturbation theory we transform the torsion components $T^a{}_{\mu\nu}$ to the purely spacetime index components $T^\alpha{}_{\mu\nu}$, which are used in the Lagrangian and find the lowest order non-vanishing term in NGR Lagrangian~\eqref{eq:LNGR}
\begin{align}\label{eq:pertL}
	\varepsilon^2 \mathcal{G}_{\alpha\beta}{}^{\mu\nu\rho\sigma}\big( \partial_\mu \uu^\alpha{}_\nu - \partial_\mu \ww^\alpha{}_\nu \big)\big( \partial_\rho \uu^\beta{}_\sigma - \partial_\rho \ww^\beta{}_\sigma \big)
	+ \mathcal{O}(\varepsilon^3)\,.
\end{align}
The expression $\mathcal{G}_{\alpha\beta}{}^{\mu\nu\rho\sigma}$ is the zeroth order of $G_{\alpha\beta}{}^{\mu\nu\rho\sigma}$, i.e., all metric components $g_{\mu\nu}$ in \eqref{eq:G} are replaced by components of the Minkowski metric $\eta_{\mu\nu}$. Observe that the Lagrangian of every teleparallel theory of gravity, which is constructed from the torsion and the tetrad alone without involving higher derivatives of the tetrad, has a lowest order term of the kind \eqref{eq:pertL}.

The field equations to lowest non-trivial order are now easily obtained from the Euler-Lagrange equations. The Lagrangian only depends on the derivative of the fundamental variables $\uu$ and $\ww$ and thus we find
\begin{align}
	0 &= \partial_\lambda\frac{\partial L}{\partial \partial_\lambda \uu^\tau{}_\kappa} \Leftrightarrow 0 = \mathcal{G}_{\tau\beta}{}^{\lambda\kappa\rho\sigma}\partial_\lambda\big( \partial_\rho \uu^\beta{}_\sigma - \partial_\rho \ww^\beta{}_\sigma \big)\,,\\
	0 &= \partial_\lambda\frac{\partial L}{\partial \partial_\lambda \ww^\tau{}_\kappa} \Leftrightarrow 
	0 =(\mathcal{G}_{\tau\beta}{}^{\lambda\kappa\rho\sigma} - \eta_{\gamma\tau}\eta^{\xi\kappa}\mathcal{G}_{\xi\beta}{}^{\lambda\gamma\rho\sigma})\partial_\lambda\big( \partial_\rho \uu^\beta{}_\sigma - \partial_\rho \ww^\beta{}_\sigma \big)\,,
\end{align}
where we use the antisymmetry of $\ww_{\mu\nu}$ in its indices to derive the second equation, or, in other words, allowed only antisymmetric variations of $\ww$; note that due to our restriction~\eqref{eqn:flatspinconn} to flat spin connections this is essentially the linearized version of the restricted variation method introduced in~\cite{Golovnev:2017dox}.
Raising the index $\tau$ the equations can be written as
\begin{align}
	0 &= \mathcal{G}^{\tau\beta\lambda\kappa\rho\sigma}\partial_\lambda\partial_\rho(\uu_{\beta\sigma} - \ww_{\beta\sigma})\,,\\
	0 &= \mathcal{G}^{[\tau|\beta\lambda|\kappa]\rho\sigma} \partial_\lambda\partial_\rho(\uu_{\beta\sigma} - \ww_{\beta\sigma})\,.
\end{align}
It is clear that these two sets of equations are not independent of each other, but the latter is the antisymmetric part of the former, a feature that has been discussed in the context of the covariant formulation of teleparallel theories of gravity \cite{Krssak:2015oua,Hohmann:2017duq}. Moreover it is clear that $\uu$ and $\ww$ are not independent variables of the theory.

To proceed we introduce the new gauge invariant (compare~\eqref{eq:pertgt}) variable $\xx_{\beta\sigma} = \uu_{\beta\sigma} - \ww_{\beta\sigma}$ which must satisfy the field equations
\begin{align}\label{eq:fieldeq}
	0 &= \mathcal{G}^{\tau\beta\lambda\kappa\rho\sigma}\partial_\lambda\partial_\rho \xx_{\beta\sigma}\,.
\end{align}
For further simplification we decompose $\xx_{\beta\sigma}$ into its symmetric and antisymmetric part $\xx_{\beta\sigma} = s_{\beta\sigma} + a_{\beta\sigma}$ which allows us to analyse the field equations further. Using this decomposition and the explicit form of $\mathcal{G}$, see \eqref{eq:G}, they take the form
\begin{align}
	0 = E^{\tau\kappa}
	& = \partial_\rho \big[ (2 c_1 - c_2 ) \partial^\rho a^{\tau\kappa} - (2 c_1 - c_2) \partial^\kappa a^{\tau\rho} + (2 c_2 + c_3)\partial^\tau a^{\rho\kappa}\big]\nonumber\\
	& + \partial_\rho \big[ (2 c_1 + c_2 ) \partial^\rho s^{\tau\kappa} - (2 c_1 + c_2) \partial^\kappa s^{\tau\rho} + c_3 \big( \eta^{\tau\kappa} (\partial^\rho s^\beta{}_\beta - \partial_\beta s^{\rho\beta}) - \eta^{\tau\rho}(\partial^\kappa s^\beta{}_\beta - \partial_\beta s^{\kappa\beta}) \big)\big]\,. \label{eqn:linvaceom}
\end{align}
These equations can further be decomposed into a symmetric and into an anti-symmetric part, which are independent and given by
\begin{align}
0
& = \partial_\rho \big[ - (2 c_1 + c_2 + c_3)\partial^{(\tau} a^{\kappa)\rho}\big]\nonumber\\
& + \partial_\rho \big[ (2 c_1 + c_2 ) \partial^\rho s^{\tau\kappa} - (2 c_1 + c_2 + c_3) \partial^{(\tau} s^{\kappa)\rho} + c_3 \big( \eta^{\tau\kappa} (\partial^\rho s^\beta{}_\beta - \partial_\lambda s^{\rho\lambda}) - \eta^{\rho(\tau}\partial^{\kappa)} s^\beta{}_\beta \big)\big]\,, \label{eq:sym}\\
0
& = \partial_\rho \big[ (2 c_1 - c_2 ) \partial^\rho a^{\tau\kappa} + (2 c_1 - 3 c_2 - c_3)\partial^{[\tau} a^{\kappa]\rho}\big] + \partial_\rho \big[ (2c_1 + c_2 + c_3) \partial^{[\tau} s^{\kappa]\rho}) \big)\big]\,.\label{eq:anti}
\end{align}
Observe that for $(2 c_1 + c_2 + c_3) = 0$ the symmetric and the antisymmetric field equations decouple. If one further demands that \eqref{eq:anti} vanishes identically, in addition $(2 c_1 - c_2 ) = 0$ and $(2 c_1 - 3 c_2 - c_3) = 0$ have to be satisfied, which implies $c_1 = - \frac{1}{4} c_3$ and $c_2 = -\frac{1}{2}c_3$. Hence for all theories, whose Lagrangian is a multiple of the TEGR Lagrangian, the antisymmetric part of the field equations is satisfied trivially, and only for those. We like to point out that linearized field equations in the case of TEGR has been studied in \cite{Obukhov:2002hy} and the fully general case, albeit in a different representation, in \cite{MHN,Kuhfuss1986}.

In the following we will deduce the propagation velocity and the polarization modes of the perturbations from these field equations.

\section{Principal polynomial and speed of propagation}\label{sec:speed}
The propagation of waves satisfying a partial differential equation is determined by the principal symbol and principal polynomial of the field equations \cite{Hoermander1, Hoermander2}. The vanishing of the principal polynomial defines the wave covectors $k$ of the propagating degrees of freedom of the theory, and thus their propagation velocity.

The principal symbol is the highest derivative term of the field equations where the partial derivatives are replaced by wave covectors $\partial \to i k$. Here this corresponds to considering the field equations in Fourier space. From \eqref{eq:fieldeq} we find
\begin{align}
	0 = \mathcal{G}^{\tau\beta\lambda\kappa\rho\sigma}k_\lambda k_\rho \hat x_{\beta\sigma} = P^{\tau\beta\kappa\sigma}(k) \hat x_{\beta\sigma}\,,
\end{align}
where $\hat x_{\beta\sigma}$ is the Fourier transform of our original field variable $x_{\beta\sigma}$ and
\begin{align}
	P^{\tau\beta\kappa\sigma}(k) = \frac{c1}{2}\eta^{\tau\beta}(\eta(k,k)\eta^{\kappa\sigma} - k^\kappa k^\sigma)
	&- \frac{c2}{4}(k^\beta k^\kappa \eta^{\sigma\tau} - k^\beta k^\tau \eta^{\kappa\sigma} + k^\sigma k^\tau \eta^{\beta\kappa} - \eta(k,k) \eta^{\beta\kappa}\eta^{\sigma\tau})\nonumber\\
	&- \frac{c3}{4}(k^\tau k^\kappa \eta^{\sigma\beta} - k^\beta k^\tau \eta^{\kappa\sigma} + k^\sigma k^\beta \eta^{\tau\kappa} - \eta(k,k) \eta^{\tau\kappa}\eta^{\sigma\beta})\,.
\end{align}
The principal polynomial $P(k)$ is given by the determinant of the principal symbol, which is interpreted as a metric on the space of fields $y^{\tau\kappa} = P^{\tau\beta\kappa\sigma}(k) \hat x_{\beta\sigma}$.

From the antisymmetry of the field equations in the indices ${}^{\lambda\kappa}$ and ${}^{\rho\sigma}$ it is immediately clear that the principal symbol is degenerate, since fields of the form $\hat x_{\beta\sigma} = k_{\sigma}V_{\beta}(k)$ solve the field equations trivially. This is a clear sign of the presence of gauge degrees of freedom in the theory. In order to derive the principal symbol we must restrict the field equations to the subspace of fields, on which they are non-degenerate. This feature is common in field theories with gauge degrees of freedom and appears also in general premetric theories of electrodynamics~\cite{Pfeifer:2016har} for example.

The field equations can be seen as a map from the space of $4 \times 4$ matrices $\hat x_{\beta\sigma}$ to its duals. To identify the subspace $\mathcal{V}$ of all $4 \times 4$ matrices on which the field equations are non-degenerate we employ the following decomposition
\begin{align}\label{eq:xdecomp}
	\hat x_{\beta\sigma} = k_\beta k_\sigma U + V_\beta k_\sigma + k_\beta W_\sigma + Q_{\beta\sigma}\,,
\end{align}
where the scalar $U$, the $1$-form components $V_\alpha$ and $W_\alpha$ and the $(0,2)$-tensor $Q_{\beta\sigma}$ satisfy the constraints
\begin{align}
k_\alpha V^\alpha = 0,\quad k_\alpha W^\alpha = 0,\quad k_\alpha Q^\alpha{}_\beta = 0\,, \quad k_\alpha Q_\beta{}^\alpha = 0\,.
\end{align}
The $4$ degrees of freedom $U$ and $V^\alpha$ cannot be dynamical, since they trivially solve the field equations. Remaining are $12$ degrees of freedom, $4-1=3$ encoded in $W_\alpha$ and $16-7=9$ in $Q_{\alpha\beta}$, which span the subspace $\mathcal{V}$. Expanding~$Q^{\tau\kappa}$ further into its symmetric traceless and antisymmetric part as well as its trace by writing $Q^{\tau\kappa} = S^{\tau\kappa} + A^{\tau\kappa} + \frac{1}{3} (\eta^{\tau\kappa} - \frac{k^\tau k^\kappa}{\eta(k,k)}) Q^\sigma{}_\sigma$, and using \eqref{eq:G}, the Fourier space field equations become
\begin{align}
0 = \hat E^{\tau\kappa}= (2 c_1 + c_2 + c_3) \eta(k,k) k^\tau W^\kappa
&+ (2 c_1 + c_2) \eta(k,k) S^{\tau\kappa} + (2c_1 - c_2) \eta(k,k) A^{\tau\kappa} \nonumber\\
&+\tfrac{1}{3} Q^\sigma{}_\sigma \eta(k,k) \big( 2 c_1 + c_2 +3 c_3 \big)\big ( \eta^{\kappa\tau} - \tfrac{k^\tau k^\kappa}{\eta(k,k)}\big) \,,
\end{align}
where we use, for the sake of readability, the notation $\eta(k,k) = \eta^{\mu\nu}k_\mu k_\nu$. To analyse them further we observe that they decompose into their contractions with $k$, their trace, their symmetric traceless and antisymmetric part
\begin{subequations}
\begin{align}
0 &= \hat E^{\tau\kappa} k_\tau k_\kappa,\quad 0 = \hat E^{\tau\kappa} k_\kappa\,, \\
0 &= \hat E^{\tau\kappa} k_\tau =(2 c_1 + c_2 + c_3) \eta(k,k)^2W^\kappa\,, \label{eq:vec}\\
0 &= \hat E^{\tau}{}_{\tau} =(2 c_1 + c_2 + 3 c_3)\eta(k,k) Q^\tau{}_\tau\,,\label{eq:tr}\\
0 &= \hat E^{[\tau\kappa]}  - \tfrac{k^{[\tau}\hat E^{|\sigma|\kappa]} k_\sigma}{\eta(k,k)} = (2c_1 - c_2) \eta(k,k) A^{\tau\kappa}\,,\label{eq:asym}\\
0 &= \hat E^{(\tau\kappa)} - k^{(\tau}\hat E^{|\sigma|\kappa)} k_\sigma - \tfrac{1}{3} \big ( \eta^{\tau\kappa} - \tfrac{k^\tau k^\kappa}{\eta(k,k)}\big)\hat E^{\sigma}{}_{\sigma}= (2 c_1 + c_2) \eta(k,k) S^{\tau\kappa}\,.\label{eq:symf}
\end{align}
\end{subequations}
The first two equations are satisfied trivially for any choice of parameters $c_1, c_2$ and $c_3$. The remaining four non-trivial field equations can be represented by a block diagonal matrix acting on a field space vector which is an element of $\mathcal{V}$
\begin{align}\label{eq:matrixf}
	\eta(k,k)
	\left(\begin{array}{cccc}
	(2 c_1 + c_2 + c_3) \eta(k,k)& 0 & 0 & 0\\
	0 & (2 c_1 + c_2 + 3 c_3)& 0 & 0\\
	0 & 0 & (2c_1 - c_2) & 0\\
	0 & 0 & 0 & (2 c_1 + c_2) \\
	\end{array}\right)
	\left(\begin{array}{c}
	W^\kappa\\
	Q^\tau{}_\tau\\
	\hat A^{\tau\kappa} \\
	\hat S^{\tau\kappa}\\
	\end{array}\right)
	=
	\left(\begin{array}{c}
	0\\
	0\\
	0 \\
	0
\end{array}\right)\,.
\end{align}
Due to their simple nature the principal polynomial is now easily obtained as determinant of the above matrix
\begin{align}\label{eq:pp}
	P(k) = (2 c_1 + c_2 + c_3)^3(2 c_1 + c_2 + 3 c_3) (2c_1 - c_2)^3(2 c_1 + c_2)^5 (\eta(k,k))^{15}\,.
\end{align}
A necessary non-trivial solution of the field equations have to satisfy, is, that their wave covectors $k$ are such that $P(k) = 0$. From the above equation \eqref{eq:pp} it is evident that only null covectors of the Minkowski metric $\eta(k,k) = 0$ realize this condition. Hence we find that for NGR theories of gravity, perturbations propagate with the speed of light determined by Maxwell electrodynamics on Minkowski spacetime.

We would like to remark that this feature can also already be seen from the decomposed Fourier space field equations~\eqref{eq:vec} to~\eqref{eq:symf}. For all field equations there can only exist non-trivial solution of the field $W^\kappa, Q^\tau{}_\tau, S^{\tau\kappa}$ or $A^{\tau\kappa}$ if and only if $\eta(k,k)$ vanishes, so all field modes in the theory are massless. For the $W^\kappa$ mode we find a double pole in its propagator, which is consistent with \cite{Kuhfuss1986,VanNieuwenhuizen:1973fi}. For $\eta(k,k)\neq 0$ the only solution of the field equations is that the fields themselves vanish identically.

\section{Newman-Penrose formalism and polarizations}\label{sec:polar}
We now focus on the polarization of gravitational waves. As we have seen in the previous section, gravitational waves in New General Relativity are described by Minkowski null waves, independently of the choice of the parameters \(c_1, c_2, c_3\). This allows us to make use of the well-known Newman-Penrose formalism~\cite{Newman:1961qr} in order to decompose the linearized field equations into components, which directly correspond to particular polarizations. We then employ the classification scheme detailed in~\cite{Eardley:1973br,Eardley:1974nw}, which characterizes the allowed polarizations of gravitational waves in a given gravity theory by a representation of the little group, which is the two-dimensional Euclidean group \(\mathrm{E}(2)\) in case of null waves. In this section we determine the \(\mathrm{E}(2)\) class of New General Relativity for all possible values of the parameters \(c_1, c_2, c_3\).

The main ingredient of the Newman-Penrose formalism is the choice of a particular complex double null basis of the tangent space. In the following, we will use the notation of~\cite{Will:1993ns} and denote the basis vectors by \(l^{\mu}, n^{\mu}, m^{\mu}, \bar{m}^{\mu}\). In terms of the canonical basis vectors of the Cartesian coordinate system they are defined as
\begin{equation}
l = \partial_0 + \partial_3\,, \quad
n = \frac{1}{2}(\partial_0 - \partial_3)\,, \quad
m = \frac{1}{\sqrt{2}}(\partial_1 + i\partial_2)\,, \quad
\bar{m} = \frac{1}{\sqrt{2}}(\partial_1 - i\partial_2)\,.
\end{equation}
We now consider a plane wave propagating in the positive \(x^3\) direction, which corresponds to a single Fourier mode. The wave covector then takes the form \(k_{\mu} = -\omega l_{\mu}\) and the symmetric and antisymmetric parts of the tetrad perturbations can be written in the form
\begin{equation}\label{eqn:zwave}
s_{\mu\nu} = S_{\mu\nu}e^{i\omega u}\,, \quad
a_{\mu\nu} = A_{\mu\nu}e^{i\omega u}\,,
\end{equation}
where we introduced the retarded time \(u = x^0 - x^3\) and the wave amplitudes are denoted \(S_{\mu\nu}\) and \(A_{\mu\nu}\).

Recall that we consider minimal coupling between gravity and matter, i.e., coupling only through the metric seen as function of the tetrad, but not through the flat spin connection. This is the usual coupling prescription for non-spinning matter, which we will henceforth assume. It follows from this choice of the matter coupling that test particles follow the geodesics of the metric, and hence the autoparallel curves of its Levi-Civita connection. The effect of a gravitational wave on an ensemble of test particles, or any other type of gravitational wave detector, such as the mirrors of an interferometer, is therefore described by the corresponding geodesic deviation equation. The observed gravitational wave signal hence depends only on the Riemann tensor derived from the Levi-Civita connection. As shown in~\cite{Eardley:1974nw}, the Riemann tensor of a plane wave is determined completely by the six so-called electric components. For the wave~\eqref{eqn:zwave}, these can be written as
\begin{gather}
\Psi_2 = -\frac{1}{6}R_{nlnl} = \frac{1}{12}\ddot{h}_{ll}\,, \quad
\Psi_3 = -\frac{1}{2}R_{nln\bar{m}} = -\frac{1}{2}\overline{R_{nlnm}} = \frac{1}{4}\ddot{h}_{l\bar{m}} = \frac{1}{4}\overline{\ddot{h}_{lm}}\,,\nonumber\\
\Psi_4 = -R_{n\bar{m}n\bar{m}} = -\overline{R_{nmnm}} = \frac{1}{2}\ddot{h}_{\bar{m}\bar{m}} = \frac{1}{2}\overline{\ddot{h}_{mm}}\,,
\quad \Phi_{22} = -R_{nmn\bar{m}} = \frac{1}{2}\ddot{h}_{m\bar{m}}\,,\label{eqn:riemcomp}
\end{gather}
where dots denote derivatives with respect to \(u\) and the metric perturbation components \(h_{\mu\nu}\) are derived from the perturbation ansatz~\eqref{eqn:pertfields} as
\begin{equation}
g_{\mu\nu} = \eta_{\mu\nu} + h_{\mu\nu} = \eta_{\mu\nu} + \varepsilon(\eta_{\mu\rho}u^{\rho}{}_{\nu} + \eta_{\nu\rho}u^{\rho}{}_{\mu}) = \eta_{\mu\nu} + 2\varepsilon s_{\mu\nu}\,.
\end{equation}
Note that they depend only on the symmetric perturbation of the tetrad, so that these are the only components whose presence or absence we must determine. We now examine which of the components~\eqref{eqn:riemcomp} may occur for gravitational waves satisfying the linearized field equations~\eqref{eqn:linvaceom}.

\begin{figure}
\includegraphics[width=0.6\textwidth]{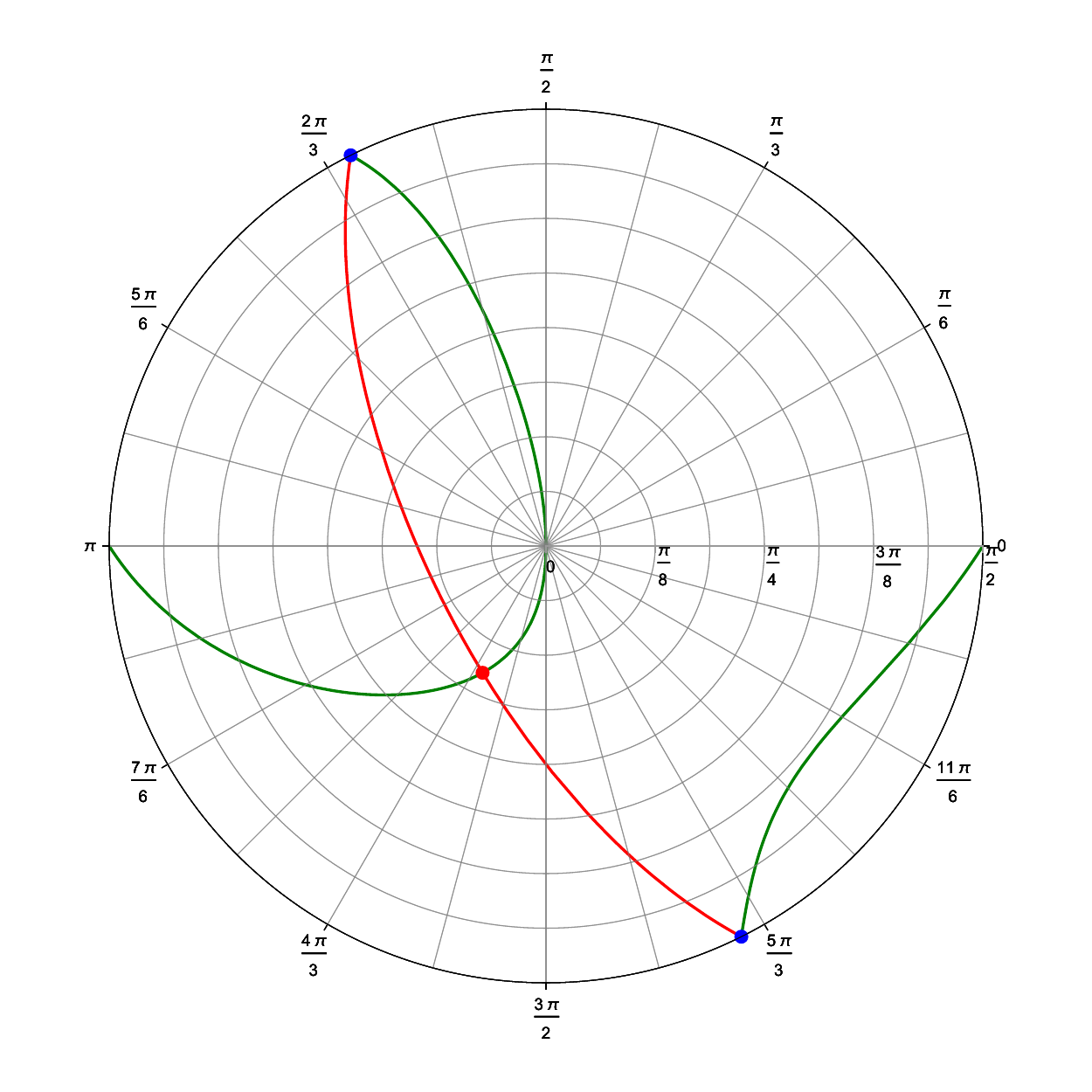}
\caption{(Color online.) Visualization of the parameter space using polar coordinates. The radial axis shows the zenith angle \(\theta\), while the (circular) polar axis shows the azimuth angle \(\phi\), following the definition~\eqref{eqn:polcoord}.
Blue Points: $2c_1 + c_2 = c_3 = 0$, class \(\mathrm{II}_6\), 6 polarizations;
green line: $2c_1 + c_2 + c_3 \neq 0, 2c_1(c_2+c_3)+c_2^2 = 0$, class \(\mathrm{III}_5\), 5 polarizations;
white area: $2c_1(c_2 + c_3) + c_2^2 \neq 0, 2c_1 + c_2 + c_3 \neq 0$, class \(\mathrm{N}_3\), 3 polarizations;
red line: $2c_1 + c_2 + c_3 = 0, c_3\neq 0$, class \(\mathrm{N}_2\), 2 polarizations.}
\label{fig:ngrpol}
\end{figure}

Inserting the wave ansatz~\eqref{eqn:zwave} and writing the gravitational Euler-Lagrange tensor \(E_{\mu\nu}\) in the Newman-Penrose basis, we find that the eight component equations
\begin{equation}
E_{ll} = E_{lm} = E_{ml} = E_{nl} = E_{mm} = E_{\bar{m}\bar{m}} = E_{l\bar{m}} = E_{\bar{m}l} = 0
\end{equation}
are satisfied identically, while the remaining eight component equations take the form
\begin{subequations}\label{eqn:np_wave}
\begin{align}
0 &= E_{nn} = (2c_1 + c_2 + c_3)\ddot{s}_{nl} + 2c_3\ddot{s}_{m\bar{m}} + (2c_1 + c_2 + c_3)\ddot{a}_{nl}\,,\label{eqn:np_nn}\\
0 &= E_{mn} = (2c_1 + c_2)\ddot{s}_{ml} + (2c_1 - c_2)\ddot{a}_{ml}\,,\label{eqn:np_mn}\\
0 &= E_{\bar{m}n} = (2c_1 + c_2)\ddot{s}_{\bar{m}l} + (2c_1 - c_2)\ddot{a}_{\bar{m}l}\,,\label{eqn:np_bmn}\\
0 &= E_{nm} = -c_3\ddot{s}_{ml} + (2c_2 + c_3)\ddot{a}_{ml}\,,\label{eqn:np_nm}\\
0 &= E_{n\bar{m}} = -c_3\ddot{s}_{\bar{m}l} + (2c_2 + c_3)\ddot{a}_{\bar{m}l}\,,\label{eqn:np_nbm}\\
0 &= E_{m\bar{m}} = E_{\bar{m}m}= -c_3\ddot{s}_{ll}\,,\label{eqn:np_mbm}\\
0 &= E_{ln} = (2c_1 + c_2)\ddot{s}_{ll}\,.\label{eqn:np_ln}
\end{align}
\end{subequations}
We now distinguish the following cases, which are also visualized in the diagram in figure~\ref{fig:ngrpol} which we explain later in this section:

\begin{itemize}
\item
\(2c_1 + c_2 = c_3 = 0\):
In this case equations~\eqref{eqn:np_mbm} and~\eqref{eqn:np_ln} are satisfied identically for arbitrary amplitudes \(S_{ll}\). For waves of this type the corresponding component \(R_{nlnl} = -6\Psi_2\) of the Riemann tensor, which describes a longitudinally polarized wave mode, is allowed to be nonzero. Following the classification detailed in~\cite{Eardley:1974nw}, they belong to the \(\mathrm{E}(2)\) class \(\mathrm{II}_6\) with six polarizations. This case corresponds to the two blue points in figure~\ref{fig:ngrpol}, which is actually a line in the three-dimensional parameter space, and hence a single point in the projected parameter space shown in the diagram, which happens to lie on the cut \(c_3 = 0\) and hence appears twice on the circular perimeter.

\item
\(2c_1(c_2 + c_3) + c_2^2 = 0\) and \(2c_1 + c_2 + c_3 \neq 0\):
It follows from the second condition that at least one of \(2c_1 + c_2\) or \(c_3\) must be nonzero. Hence, either equation~\eqref{eqn:np_mbm} or~\eqref{eqn:np_ln} imposes the condition \(S_{ll} = 0\), so that there is no longitudinal mode \(\Psi_2\). The first condition is equivalent to a vanishing determinant of the matrix
\begin{equation}
\begin{pmatrix}
2c_1 + c_2 & 2c_1 - c_2\\
-c_3 & 2c_2 + c_3
\end{pmatrix}\,,
\end{equation}
so that the equations~\eqref{eqn:np_mn}, \eqref{eqn:np_bmn}, \eqref{eqn:np_nm} and~\eqref{eqn:np_nbm} allow for non-vanishing solutions. This further implies that the two columns of this matrix are linearly dependent, and hence proportional to each other. However, from the second condition further follows that neither column vanishes. Hence, at least one of the pairs~\eqref{eqn:np_mn}, \eqref{eqn:np_bmn} and~\eqref{eqn:np_nm}, \eqref{eqn:np_nbm} of equations must be non-trivial, with the coefficients of both the symmetric and the antisymmetric tetrad perturbation non-vanishing. Hence, non-vanishing solutions of these equations have both symmetric and antisymmetric contributions, and therefore in particular non-vanishing \(S_{lm}\) and \(S_{l\bar{m}}\); however, recall that the antisymmetric part does not contribute to the geodesic deviation equation, and so we do not discuss it here. It then follows that \(R_{nln\bar{m}} = -2\Psi_3\), whose complex components describe two vector polarizations, is allowed to be nonzero. Waves of this type belong to the \(\mathrm{E}(2)\) class \(\mathrm{III}_5\) encompassing five polarizations. This case is represented by the green line in figure~\ref{fig:ngrpol}.

\item
\(2c_1(c_2 + c_3) + c_2^2 \neq 0\) and \(2c_1 + c_2 + c_3 \neq 0\):
In this case the only linearized field equation which allows for non-vanishing solutions is equation~\eqref{eqn:np_nn}. Here the only relevant component for the geodesic deviation is \(S_{m\bar{m}}\), so that we can neglect the other terms. This component is allowed to be non-vanishing, and hence allows a non-vanishing component \(R_{nmn\bar{m}} = -\Phi_{22}\) of the Riemann tensor. The corresponding scalar wave mode is called the breathing mode. The remaining equations impose the condition \(\Psi_2 = \Psi_3 = 0\), so that the longitudinal and vector modes are prohibited. This wave has the \(\mathrm{E}(2)\) class \(\mathrm{N}_3\), and thus three polarizations. Almost all points of the parameter space, shown in white in figure~\ref{fig:ngrpol}, belong to this class.

\item
\(2c_1 + c_2 + c_3 = 0\) and \(c_3 \neq 0\):
It follows immediately from equation~\eqref{eqn:np_mbm} that \(S_{ll} = 0\), so that the longitudinal mode \(\Psi_2\) is prohibited. Taking the sum of the pairs~\eqref{eqn:np_mn}, \eqref{eqn:np_bmn} and~\eqref{eqn:np_nm}, \eqref{eqn:np_nbm} of equations and replacing \(c_2\) by \(-2c_1 - c_3\) one further finds that also \(S_{lm} = S_{l\bar{m}} = 0\), and hence also the vector modes \(\Psi_3\) must vanish. Finally, equation~\eqref{eqn:np_nn} imposes the condition \(S_{m\bar{m}} = 0\), so that also the breathing mode \(\Phi_{22}\) is prohibited. It thus follows that the only unrestricted electric components of the Riemann tensor are \(R_{nmnm} = -\bar{\Psi}_4\) and its complex conjugate, corresponding to two tensor modes. The \(\mathrm{E}(2)\) class of this wave is \(\mathrm{N}_2\), with two polarizations. This case is shown as a red line in figure~\ref{fig:ngrpol}. Note in particular that TEGR, marked as a red point, belongs to this class, as one would expect. This subclass corresponds to the so-called one parameter family of teleparallel models and has received particular attention in previous studies~\cite{MHN}. It has been argued that this condition is necessary to avoid ghosts~\cite{VanNieuwenhuizen:1973fi,Kuhfuss1986}. However, we will not address the question of ghosts in this article, and leave this discussion for a separate study.
\end{itemize}

We have visualized the aforementioned cases in figure~\ref{fig:ngrpol}, which we constructed as follows. We first made use of our assumption that at least one of the parameters \(c_1, c_2, c_3\) is non-vanishing and introduced normalized parameters
\begin{equation}
\tilde{c}_i = \frac{c_i}{\sqrt{c_1^2 + c_2^2 + c_3^2}}
\end{equation}
for \(i = 1, 2, 3\). One easily checks that the \(\mathrm{E}(2)\) classes we found only depend on these normalized parameters. We then introduced polar coordinates \((\theta, \phi)\) on the unit sphere to express the parameters as
\begin{equation}\label{eqn:polcoord}
\tilde{c}_1 = \sin\theta\cos\phi\,, \quad
\tilde{c}_2 = \sin\theta\sin\phi\,, \quad
\tilde{c}_3 = \cos\theta\,.
\end{equation}
Since the \(\mathrm{E}(2)\) class is the same for antipodal points on the parameter sphere, we restrict ourselves to the hemisphere \(\tilde{c}_3 \geq 0\), and hence \(0 \leq \theta \leq \frac{\pi}{2}\); this is equivalent to identifying antipodal points on the sphere and working with the projective sphere instead, provided that we also identify antipodal points on the equator \(\tilde{c}_3 = 0\). We then considered \((\theta, \phi)\) as polar coordinates on the plane in order to draw the diagram shown in figure~\ref{fig:ngrpol}. Note that antipodal points on the perimeter, such as the two blue points, are identified with each other, since they describe the same class of theories.

This concludes our discussion of gravitational wave polarizations. We have seen that depending on the parameters \(c_1, c_2, c_3\) we obtain the \(\mathrm{E}_2\) class \(\mathrm{II}_6\), \(\mathrm{III}_5\), \(\mathrm{N}_3\) or \(\mathrm{N}_2\), with \(\mathrm{N}_3\) filling most of the parameter space. We have also seen that there exists a family of theories besides TEGR which is of class \(\mathrm{N}_2\) and thus exhibits the same two tensor modes as in general relativity. Theories in this class therefore cannot be distinguished from general relativity by observing the polarizations of gravitational waves alone.

\section{Conclusion}\label{sec:conclusion}
We studied the propagation of gravitational waves in the most general class of teleparallel gravity theories whose action is quadratic in the torsion tensor, known as new general relativity. The wave we considered is modeled as a linear perturbation of a diagonal tetrad corresponding to a Minkowski background metric. We derived the principal polynomial of the linearized field equations and found that gravitational waves propagate at the speed of light, i.e., their wave covector must be given by a null vector of the Minkowski background. Further, we made use of the Newman-Penrose formalism to derive the possible polarizations of gravitational waves. Our results show that the two tensor polarizations, which are present also in general relativity, are allowed for the whole class of theories we considered, while additional modes - two vector modes and up to two scalar modes - may be present for particular models within this class. We found that the teleparallel equivalent of general relativity is not the unique theory exhibiting exactly two polarizations, but there is a one-parameter family of theories with the same property. It thus follows that observations of gravitational wave polarizations may only give partial results on the parameter space of these theories.

We remark that although we restricted our analysis to theories whose action is quadratic in the torsion tensor, our results are valid for a significantly larger class of theories. This is due to the fact that the torsion is linear in the tetrad perturbations, so that the action is already quadratic in the perturbations. Hence, any higher order correction terms would have no influence on the linearized field equations. This observation agrees with previous results that there are no additional gravitational polarizations in \(f(T)\) gravity compared to general relativity~\cite{Bamba:2013ooa}, since up to the required perturbation order the Lagrangian can be approximated as \(f(T) = f(0) + f'(0)T + \mathcal{O}(T^2)\), which is equivalent to general relativity with a cosmological constant. An extension to the class of theories discussed in~\cite{Bahamonde:2017wwk} is shown in~\cite{Hohmann:2018xnb}.

Although higher order terms in the action do not influence the linear perturbations around a Minkowski background, they certainly have an influence on the cosmological dynamics of the theory, and therefore on the expansion history of the Universe. This modified expansion history might thus also leave an imprint on the observed gravitational waves propagating in a cosmological background. An interesting extension of our work would be to study gravitational waves as a perturbation to a tetrad corresponding to a Friedmann-Lemaitre-Robertson-Walker metric, taking into account modifications of the background dynamics arising from higher order torsion terms. Note that such modifications do not show up in the quadratic action we considered in this article, since all terms in the gravitational action become proportional to the square of the Hubble parameter in the case of cosmological symmetry, and so the action reduces to the teleparallel equivalent of general relativity, up to a constant factor.

Another possible class of extensions is to consider additional fields non-minimally coupled to torsion and to study their influence both on the speed and the polarization of gravitational waves. A canonical example is given by scalar-torsion theories~\cite{Hohmann:2018rwf,Hohmann:2018vle,Hohmann:2018dqh,Hohmann:2018ijr} constructed from the TEGR torsion scalar and an additional scalar field, where one would expect the presence of an additional scalar mode compared to general relativity as it is also the case for scalar-curvature gravity. These theories can be extended by replacing the TEGR torsion scalar with the NGR torsion scalar which defined the Lagrangian considered in this article.

\begin{acknowledgments}
The authors would like to thank Jackson Levi Said for hospitality at the University of Malta, Tomi Koivisto for helpful comments and remarks, and the anonymous referee for pointing us to the references \cite{MHN,Kuhfuss1986}. They were supported by the Estonian Research Council through the Institutional Research Funding project IUT02-27 and the Personal Research Funding project PUT790 (start-up project), as well as by the European Regional Development Fund through the Center of Excellence TK133 ``The Dark Side of the Universe''. UU acknowledges mobility support from the European Regional Development Fund through Dora Plus.
\end{acknowledgments}

\bibliographystyle{utphys}
\bibliography{wavesTP}
\end{document}